\documentclass[sigplan,nonacm]{acmart}
\usepackage[utf8]{inputenc}
\usepackage{href-ul}
\usepackage{booktabs}
\usepackage{natbib}
\usepackage{bibcop}
\usepackage{pgfplots}
\usepackage{comment}
\usepackage{subcaption}
\usepackage{amsmath}
\usepackage{listings}
\usepackage[noframes,sf,bold]{ffcode}
\usepackage{tabularx}
\usepackage{paralist}
\usepackage[off]{to-be-determined}
\usepackage{float}
\usepackage{multicol}
\usepackage{wrapfig}
\usepackage[capitalize]{cleveref}

\makeatletter\newcommand\footnoteref[1]{\protected@xdef\@thefnmark{\ref{#1}}\@footnotemark}\makeatother

\title{Heap vs. Stack: Analyzing Memory Allocations in C and C++ Open Source Software}

\author{anonymous}

\author{Eugene Darashkevich}
\orcid{0000-0002-1655-2331}
\email{eugene.darashkevich@gmail.com}
\affiliation{Yandex\country{Russia}\city{Moscow}}

\author{Roman Rusyaev}
\orcid{0009-0008-7108-5562}
\email{rusyaev.rm@gmail.com}
\affiliation{\institution{Huawei}\country{Russia}\city{Moscow}}

\author{Roman Korostinskiy}
\orcid{0009-0008-7950-3095}
\email{korostinskii.r@phystech.edu}
\affiliation{\institution{MIPT}\country{Russia}\city{Moscow}}

\author{Yegor Bugayenko}
\orcid{0000-0001-6370-0678}
\email{yegor256@gmail.com}
\affiliation{\institution{Huawei}\country{Russia}\city{Moscow}}

\begin{document}

\begin{abstract}
In C++, objects can be allocated in static memory, on the stack, or on the heap---the latter being significantly more performance-costly than the former options. We hypothesized that programmers, particularly those involved in widely-used open-source projects, would be conscious of these performance costs and consequently avoid heap allocations. To test this hypothesis, we compiled and executed 797 automated tests across 13 C and 10 C++ open GitHub projects, measuring their heap allocations with Valgrind and stack allocations using DynamoRIO instrumentation. Our findings showed a wide variation in heap allocations, ranging from 0 to 99\% with an average of 9.26\%. We also found that C++ programs use heap less frequently than C programs. Contrary to our initial intuition, this suggests that heap allocations are actively employed in both C and C++ programs. Determining the prevalence of objects in these allocations remains a topic for future research.
\end{abstract}

\maketitle

\section{Introduction}\label{sec:intro}

\tbd{It is not clear whether the shared criteria exclusively resulted in the selected 10 C++ and 13 C projects or the authors somehow cherry picked the projects. For example, did the author sort all the projects with 1000+ star that comply with the other two criteria based on the starts (or other metrics) and selected the top 10 C++ and 13 C projects? If so, why only 10 and 13 projects? It seems an arbitrary number.}

\tbd{ what level of optimization was used during the compilation? This can affect the final assembly code generated by the compiler, and thus, affects the number (and the way of) of stack allocations.}

\tbd{Why did not the author use some well-established tools such as PIN for the instrumentation and memory allocation
logging?}

\tbd{While I understand providing input for each of 700+ test cases can be tedious, there are so many ways to automatically feed the input to the binary, e.g., fuzzing, config files, etc. It is not clear to me why the authors excluded binaries that require input?}

\tbd{This paper barely reports an engineering effort on using two memory tracing tools, i.e., Valgrind and DynamoRIO, to log the memory allocations, and fails to compare the results with SOTA.}

\tbd{There is of course nuisance in the topic, but it is hard for any reader to keep track of what the arguments are; see the
following sentences from different parts of the paper (abstract and intro seem to contradict the results and conclusion).
Abstract -- "We also found that C++ programs use heap less frequently than C programs."
Introduction -- "It may be assumed that C++ programmers understand the negative performance effect of on-heap
allocations and avoid them."
Results -- "This histogram demonstrates that the allocation patterns between C and C++ executables are quite
similar."
Conclusion -- "Contrary to the expectation that programmers might avoid heap allocations due to their high
performance costs, our results indicate that heap allocations are extensively used in both C and C++ programs."}

C++ enables programmers to encapsulate code and data within objects~\citep{cpp}, which are data structures situated in static memory, on the stack, or in the heap. While static and stack allocations are relatively efficient, heap allocation using |new| and |delete| operators is far more performance-costly and is, next to virtual tables~\citep{driesen1996direct,aigner1996eliminating}, one of two main inefficiencies of object-oriented programming~\citep{budimlic1999cost}.

It may be assumed that C++ programmers understand the negative performance effect of on-heap allocations and avoid them. If this is true, developers of compilers don't need to invest efforts into making on-heap allocations faster or reducing the number of them in the code. However, if this is not true, optimization of on-heap object allocations may look like a profitable endeavor. In order to help compiler developers get a clear picture of how often programmers allocate objects on heap, in this study we asked the following research question: how often are objects allocated on the stack versus the heap in open-source C++ projects? 

In order to answer this question, we took 23 public C/C++ GitHub repositories and compiled their automated tests into 797 executable binaries. Then, we instrumented them with the help of DynamoRIO and executed with Valgrind, thus counting the number of stack and on-heap allocations in runtime. Our findings indicate that the ratio of heap allocations vs. all allocations differs significantly between tests, ranging over the 0--99\% interval. Our measurements also demonstrate that C++ binaries allocate on heap less frequently than C binaries. Thus, it is reasonable to confirm that optimizations of on-heap allocations of objects in C/C++ programs may, in some cases, significantly improve their performance.

The rest of this paper is organized as follows: 
\cref{sec:method} elaborates on our methodology and data collection process. \Cref{sec:result} presents the empirical results, followed by a discussion and limitations in \cref{sec:discussion}. We explore related work in \cref{sec:related} and conclude in \cref{sec:conclusion}.

\section{Method}\label{sec:method}

First, we selected some open-source repositories on GitHub, adhering to the following criteria:
\begin{inparaenum}[1)]
\item they have over one thousand stars,
\item their automated tests or benchmarks can be compiled into executable binaries that do not require any input, and
\item they have an automatic build system (preferably CMake)
\end{inparaenum}

Second, we obtained the number of allocations and the number of allocated bytes for the heap using Valgrind, which seemed to be the simplest and most reliable strategy. We executed each program using Valgrind and extracted the relevant information from the ``HEAP SUMMARY'' part of its report.

Third, we counted stack allocations by monitoring modifications to the stack pointer. The stack pointer (\ff{rsp}) is integral to stack memory allocation and deallocation. Common x86-64 assembly instructions used for stack space allocation include:
\begin{itemize}
\item ``\ff{sub rsp,N}'': This instruction subtracts a constant value N from the stack pointer (\ff{rsp}) to allocate space on the stack.
\item ``\ff{push rax}'': This instruction pushes the contents of a register (such as \ff{rax}) onto the stack, effectively allocating space.
\item ``\ff{call foo}'': Push the current value of the instruction pointer (\ff{rip}) onto the stack, also decrementing the stack pointer by 8 bytes.
\end{itemize}
Hence, counting stack allocations involved tallying \ff{sub} instructions that modify the stack pointer (\ff{rsp}), \ff{push} instructions that allocate stack memory, and \ff{call} instructions that push return addresses onto the stack. 
In order to count real-time instruction, we created a profiler with the help of DynamoRIO\footnote{\url{https://github.com/DynamoRIO/dynamorio}}, that enabled precise instruction counting.

Finally, we determined the proportion of heap and stack allocations out of all allocations . These values are
given by the equations: $Q_H = H/(H + S)$, $Q_S = S/(H + S) = 1 - Q_H$. In these equation, \(H\) denotes the total number of heap object allocations, and \(S\) represents the total number of stack object allocations.

It is worth noting that we have not analyzed the source code of these executable files. Later in the article, we will discuss the limitations that are imposed on our method because of this.

\section{Results}\label{sec:result}

First, we selected 10 C++ and 13 C open\tbd{-source} repositories in GitHub. Then, we compiled the source code of all automated tests in all repositories and obtained 960 command-line executable binaries. \Cref{table:repos} shows all repositories selected together with their recent commit hashes, the number of executable binaries, and the primary programming languages used in them. We excluded 163 executables from the experiment because:
\begin{inparaenum}[1)]
    \item 73 of them did nothing; \tbd{It is not clear what does "nothing" mean here? Did the program fail? Did it end without printing something on the command prompt?}
    \item 48 of them, despite being tests, required input;
    \item 39 of them required additional dependencies (e.g. external libraries);
    \item 2 of them produced too many Valgrind errors;
    \item 1 of them yielded a non-null return code.
\end{inparaenum}

\begin{table*}
\caption{The table lists C/C++ GitHub repositories selected for the analysis, the Git commit on which the data was collected (from GitHub), and the number of suitable executables vs. the number of all compiled executables.}
\label{table:repos}
\setlength\extrarowheight{-3pt}
\begin{tabularx}{\linewidth}{>{\ttfamily}Xl>{\ttfamily}l>{\ttfamily}r>{\ttfamily}r>{\ttfamily\arraybackslash}r}
\toprule
  & 
  & 
  & \multicolumn{2}{c}{{\rmfamily Executables}}
  & 
  \\
{\rmfamily Repository} 
  & Lang 
  & {\rmfamily Commit}
  & {\rmfamily Runnable}
  & {\rmfamily Total}
  & {\rmfamily Avg $Q_H$}
  \\
\midrule
DaveGamble/cJSON 
  & C 
  & 87d8f096 
  & 19
  & 19
  & 3.67
  \\ 
commonmark/cmark 
  & C 
  & 2632fdce 
  & 1 
  & 1
  & 2.66
  \\ 
srdja/Collections-C 
  & C 
  & 6b6ec211 
  & 25
  & 25
  & 4.28
  \\ 
fastfetch-cli/fastfetch 
  & C
  & 67694107 
  & 2 
  & 2
  & 1.01
  \\ 
json-c/json-c 
  & C
  & 0051f2db 
  & 25
  & 27
  & 33.90
  \\ 
libsndfile/libsndfile 
  & C
  & c81375f0
  & 22
  & 36
  & 0.98
  \\ 
microsoft/mimalloc 
  & C
  & 4e50d671
  & 1
  & 3
  & 20.52
  \\ 
nanomsg/nanomsg 
  & C 
  & fc3f684a
  & 43
  & 50
  & 17.15
  \\ 
davidmoreno/onion 
  & C
  & de8ea938
  & 16
  & 17
  & 34.46
  \\ 
openssl/openssl 
  & C
  & 6594baf6 
  & 183
  & 295
  & 23.94
  \\ 
aws/s2n-tls 
  & C
  & b3ee9af0 
  & 259
  & 259
  & 10.48
  \\ 
kuba--/zip 
  & C
  & 6f2116d7 
  & 8
  & 8
  & 3.51
  \\ 
HardySimpson/zlog
  & C
  & 7fe61ca6
  & 2
  & 27
  & 2.83
  \\ 
google/benchmark 
  & C++ 
  & bc946b91 
  & 40
  & 40
  & 4.44
  \\ 
fmtlib/fmt 
  & C++
  & f4b256c6
  & 20
  & 20
  & 2.85
  \\ 
google/glog 
  & C++
  & 31429d85
  & 13
  & 13
  & 0.21
  \\ 
nlohmann/json 
  & C++
  & c883fb0f
  & 75
  & 75
  & 0.24
  \\ 
google/leveldb 
  & C++
  & 068d5ee1
  & 3
  & 3
  & 0.11
  \\ 
Neargye/magic\_enum 
  & C++
  & f34f967c
  & 22
  & 22
  & 0.30
  \\ 
Alinshans/MyTinySTL 
  & C++
  & acc07e02
  & 1
  & 1
  & 7.30
  \\
BYVoid/OpenCC 
  & C++
  & e5d6c5f1
  & 14
  & 14
  & 3.53
  \\ 
gabime/spdlog
  & C++
  & a2b42620
  & 2
  & 2
  & 8.88
  \\ 
jbeder/yaml-cpp 
  & C++
  & 76dc6715
  & 1
  & 1
  & 0.03
  \\
\bottomrule
\end{tabularx}
\end{table*}

We ended up with 797 binaries that we could execute without providing any input. For these tests, we employed a machine running Ubuntu 22.04, equipped with an Intel Core i5-7200U processor and 12GB of RAM. All binaries were compiled using GCC~11.4.0 in the "\texttt{RELEASE}" build configuration.

Second, we executed all binaries and collected necessary data using DynamoRIO and Valgrind (wired together with our custom scripts). We executed each binary only once and did not analyze their source code.

\begin{figure*}
\begin{subfigure}[t]{.45\textwidth}
\includegraphics[width=\columnwidth]{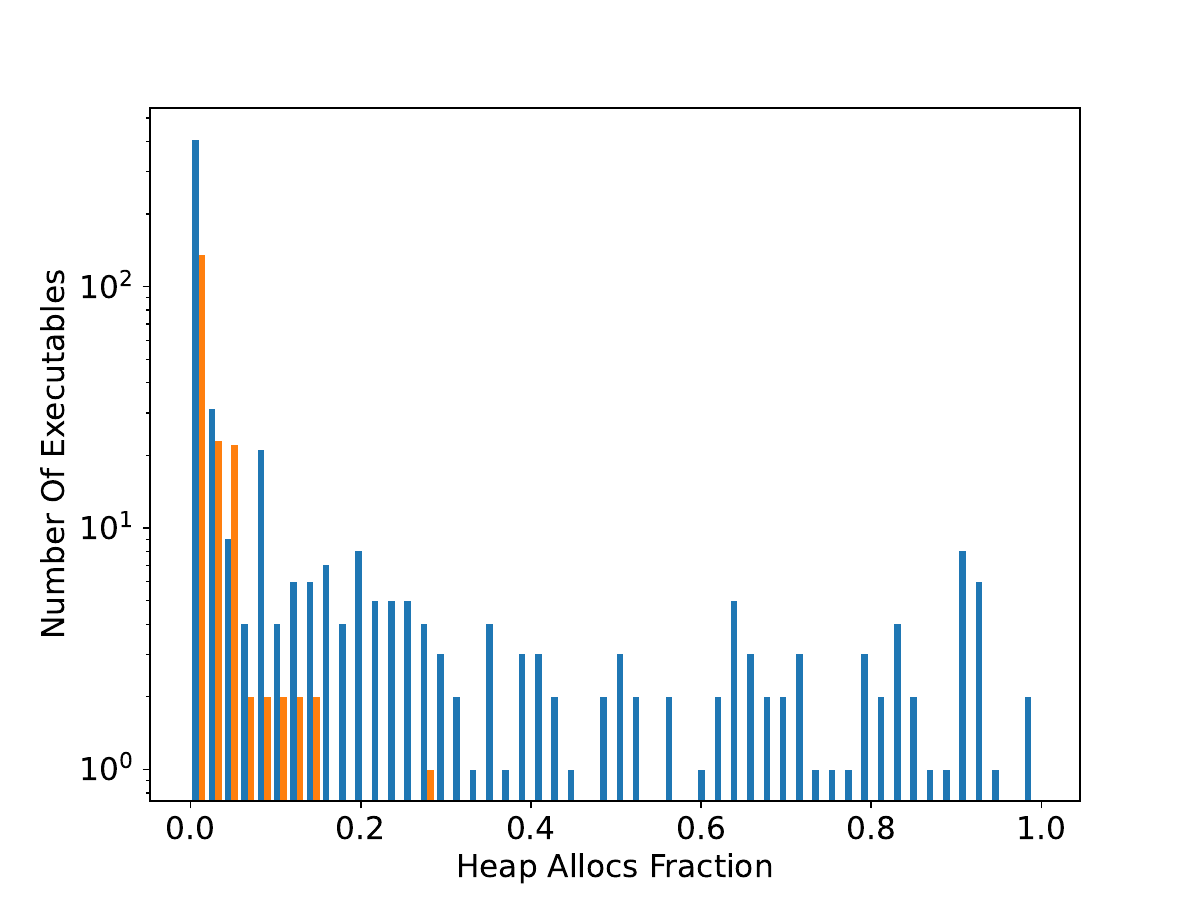}
\caption{The distribution of C (blue) and C++ (orange) executable files (separately), by the fraction of heap allocation relative to the total allocations (on a logarithmic scale).}
\label{fig:hist-c-cpp}
\end{subfigure}
\hfill
\begin{subfigure}[t]{.45\textwidth}
\includegraphics[width=\columnwidth]{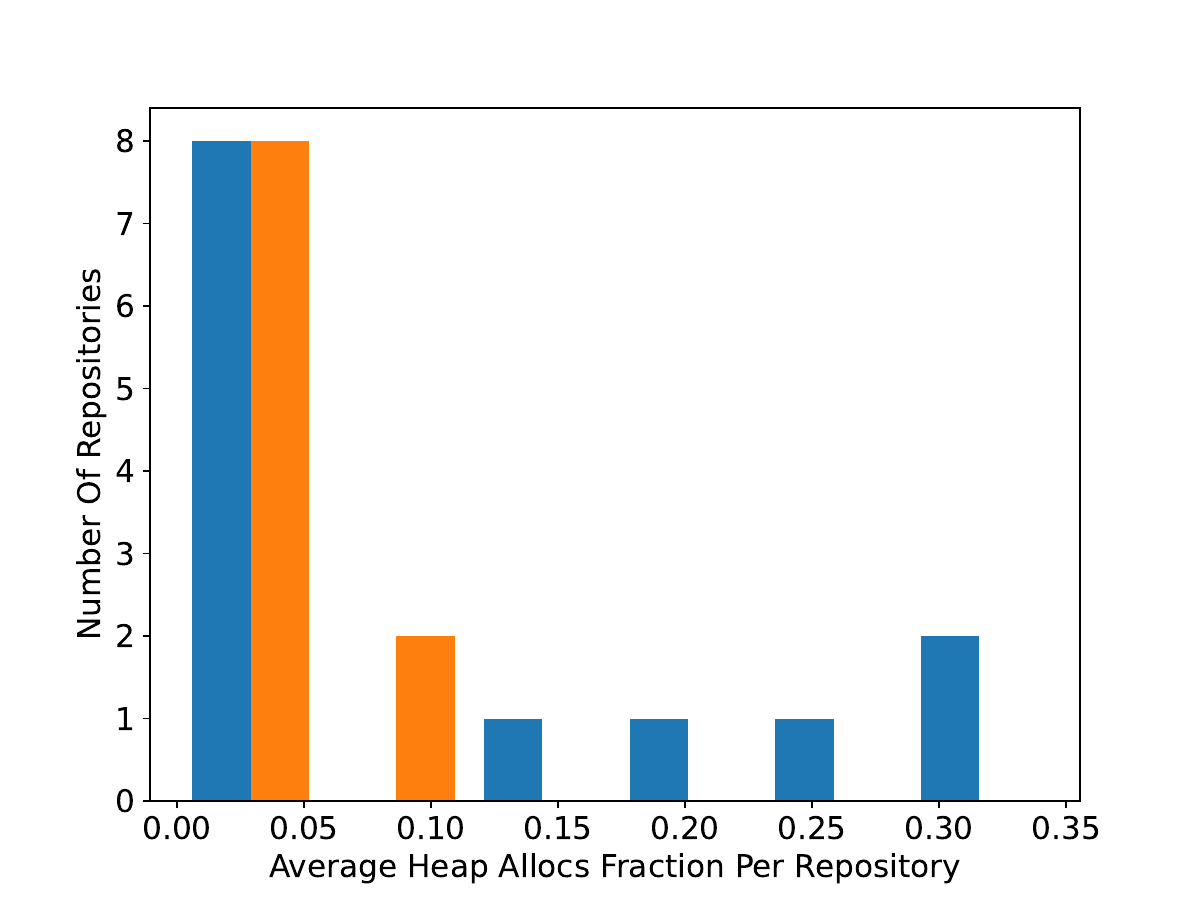}
\caption{The distribution of C (blue) and C++ (orange) repositories by the average fraction of heap allocation ($Q_H$).}
\label{fig:hist-repos-c}
\end{subfigure}
\caption{Distribution of executables by the value of the fraction of heap allocations.}
\end{figure*}

Third, we collected metrics and built plots. The main metric we wanted to know is what fraction of all allocations occur on the heap ($Q_H$). The distribution of C and C++ executables by the fraction of heap allocations is shown in the form of histogram in \cref{fig:hist-c-cpp}. This histogram demonstrates that the allocation patterns between C and C++ executables are quite similar. However, the distribution appears too chaotic to draw any far-reaching conclusions.

We have also built histograms for repositories. \cref{fig:hist-repos-c} shows the distribution of C and C++ open-source repositories which we have selected.

\begin{figure*}
\begin{subfigure}[t]{.45\textwidth}
\includegraphics[width=\columnwidth]{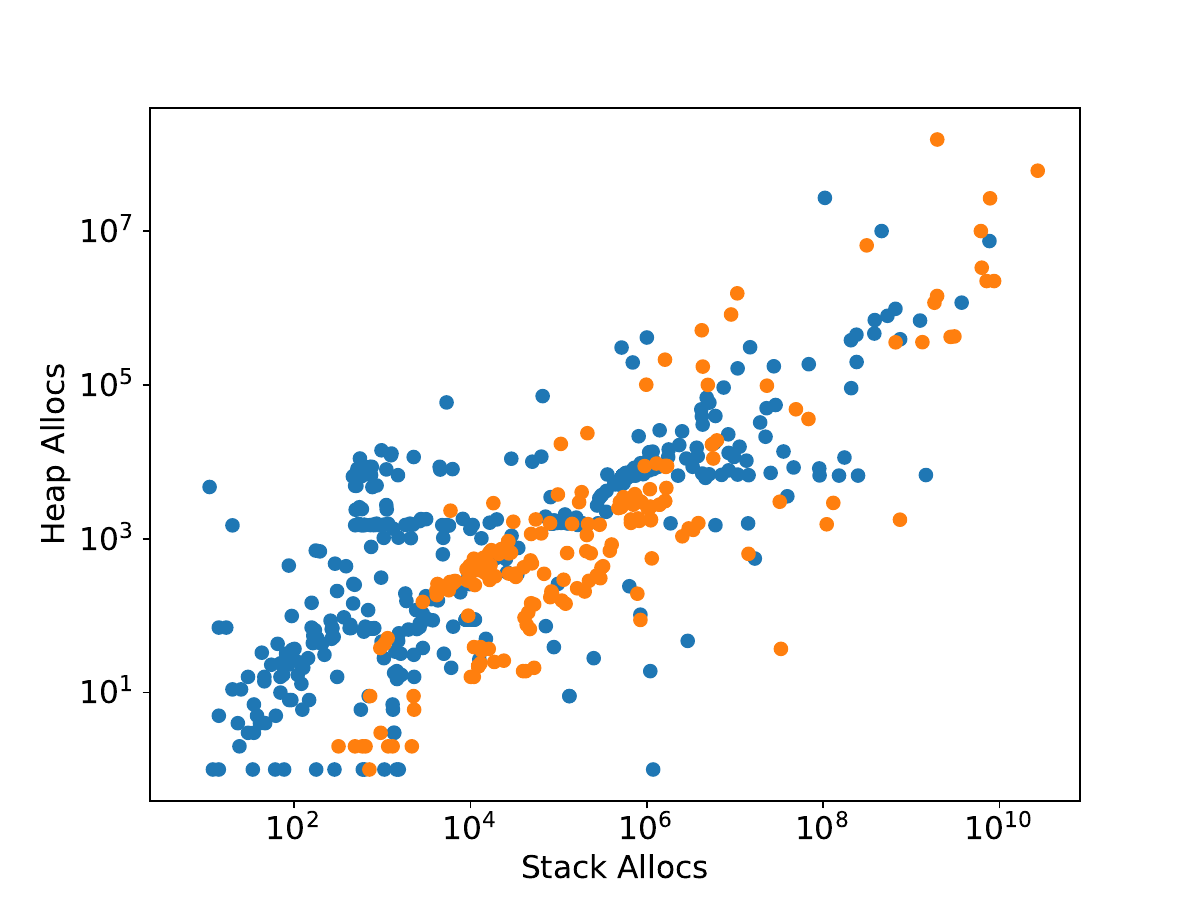}
\caption{The plot shows relationship between stack and heap allocations amount in C (blue) and C++ (orange) executables from various open source repositories. The axes are shown on a logarithmic scale.}
\label{fig:stack-allocs-heap-allocs-plot}
\end{subfigure}
\hfill
\begin{subfigure}[t]{.45\textwidth}
\includegraphics[width=\columnwidth]{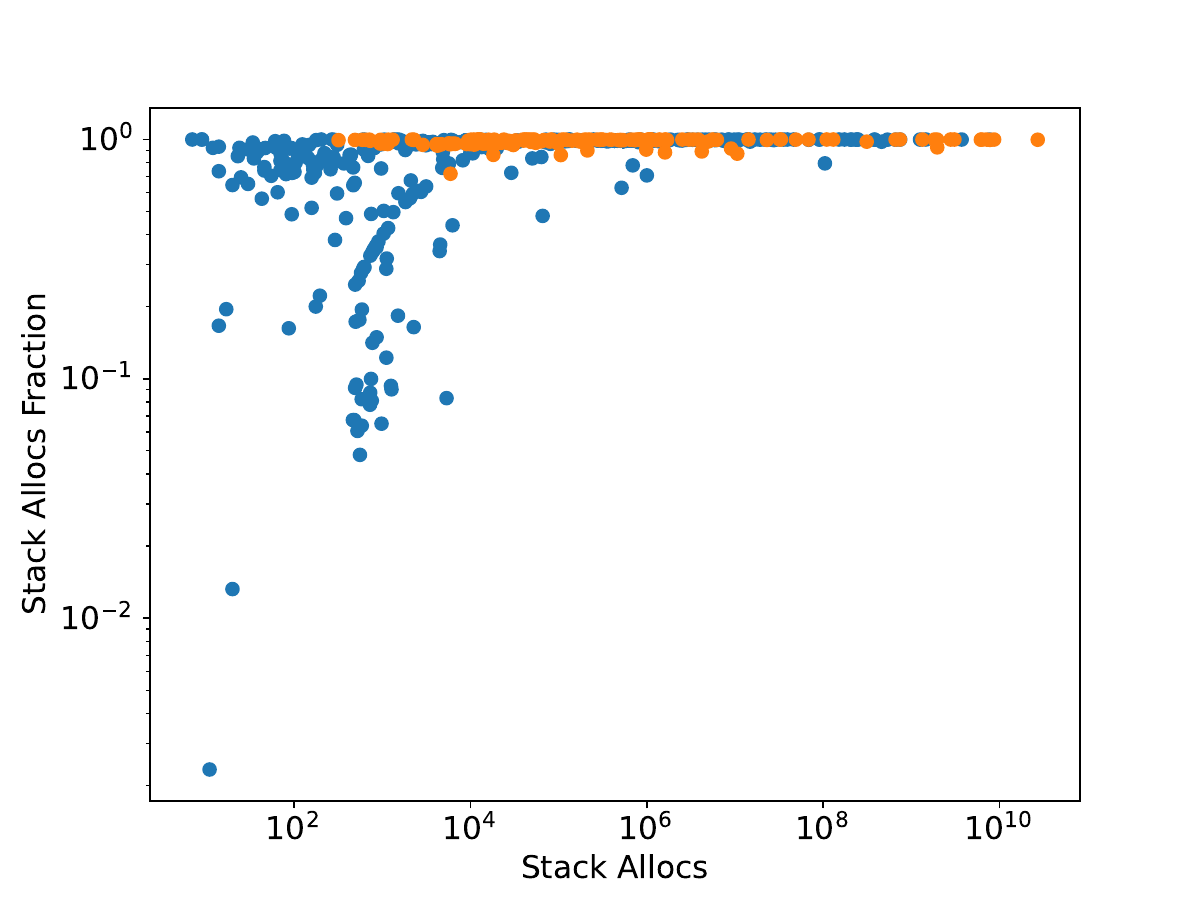}
\caption{The plot shows relationship between stack allocations amount and stack allocations fraction in C (blue) and C++ (orange) executables from various open source repositories. The axes are shown on a logarithmic scale.}
\label{fig:stack-frac-stack-allocs-plot}
\end{subfigure}
\hfill
\begin{subfigure}[t]{.45\textwidth}
\includegraphics[width=\columnwidth]{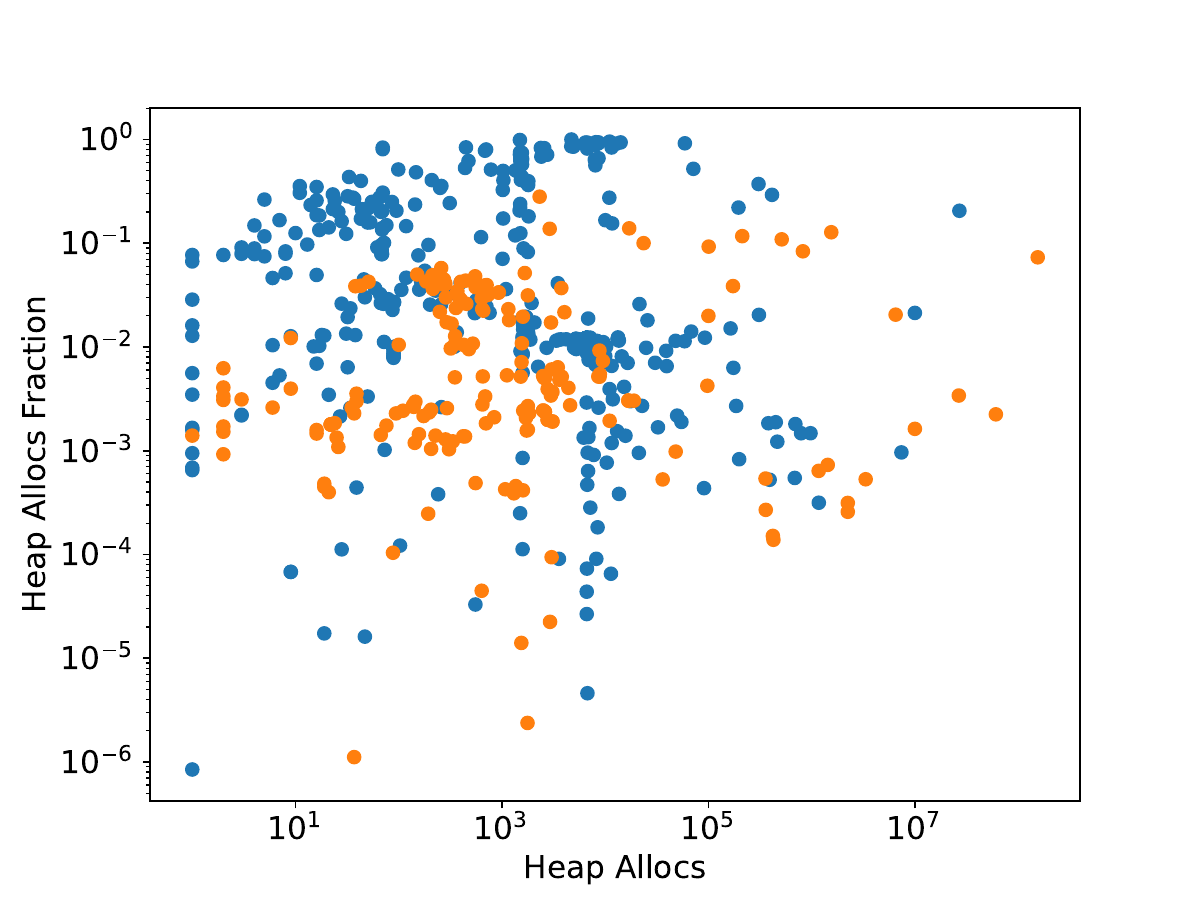}
\caption{The plot shows relationship between heap allocations amount and heap allocations fraction in C (blue) and C++ (orange) executables from various open source repositories. The axes are shown on a logarithmic scale.}
\label{fig:heap-fracs-heap-allocs-plot}
\end{subfigure}
\hfill
\begin{subfigure}[t]{.45\textwidth}
\includegraphics[width=\columnwidth]{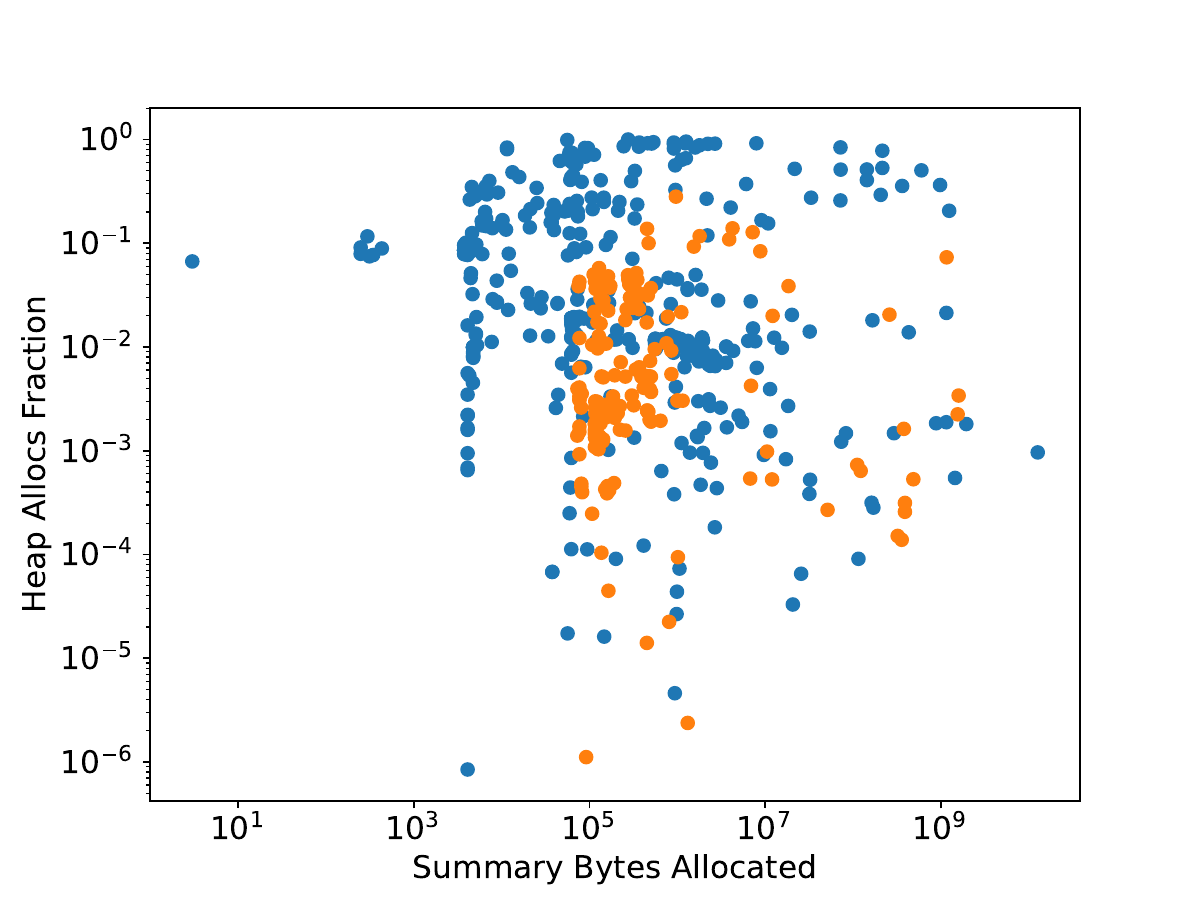}
\caption{The plot shows relationship between summary bytes allocated on heap and heap allocations fraction in C (blue) and C++ (orange) executables from various open source repositories. The axes are shown on a logarithmic scale.}
\label{fig:heap-allocs-heap-size-plot}
\end{subfigure}
\caption{Visualized relationships between certain characteristics of executables and their runtime results.}
\end{figure*}

In order to test the relationship between the different data obtained from executables we built plots from \cref{fig:stack-allocs-heap-allocs-plot}, \cref{fig:stack-frac-stack-allocs-plot}, \cref{fig:heap-fracs-heap-allocs-plot}, and \cref{fig:heap-allocs-heap-size-plot}. All plots are presented on a logarithmic scale due to the significant scale difference in the data. In \cref{fig:stack-allocs-heap-allocs-plot}, we depict the relationship between stack ($S$) and heap ($H$) allocations in C and C++ executables. \cref{fig:stack-frac-stack-allocs-plot} illustrates the correlation between the quantity of stack allocations ($S$) and the stack allocation fraction ($Q_S$) for the analyzed executables. Similarly, \cref{fig:heap-fracs-heap-allocs-plot} explores the analogous relationship for heap allocations ($H$ and $Q_H$). Finally, \cref{fig:heap-allocs-heap-size-plot} demonstrates the relationship between the number of bytes allocated on the heap and the heap allocation fraction ($Q_H$).

\section{Discussion}\label{sec:discussion}

\textbf{Why do programmers allocate objects on the heap while the stack is much cheaper?}
In \citeyear{10.1007/11535409_1}, Bjarne Stroustrup concluded the following regarding C++: "In general, C++ implementations obey the zero-overhead principle: What you don't use, you don't pay for. What you do use, you couldn't hand code any better"~\citep{10.1007/11535409_1}. This is indeed true for object allocations, because a C++ object is essentially a C |struct|, which can also be allocated either on the stack or on the heap. The motivation for placing a |struct| on the heap is the same as the reasoning for placing an object on the heap: either when the link to the object must live longer than the function that creates it, or when the size of the object is not known at compile time. There is yet another reason, though, why programmers may allocate on the heap: they are not aware of the costs. In our study, we didn't analyze whether every allocation on the heap is motivated, but we suspect that a decent amount of objects may be placed on the heap by mistake. \tbd{How did the authors come to this speculation or a guess? I was expecting to see at least a shallow empirical study on the root cause of the reported results.} This may be a subject for future research: to analyze how many objects are placed on the heap when they could live on the stack instead.

\textbf{Why did we count allocations dynamically instead of using static analysis?}
We counted heap and stack allocations for each executable binary file dynamically: we executed them. We chose this method over static analysis because it provides a more accurate count of actual allocations. A static approach would necessitate making assumptions about program execution, yielding less precise results. However, static analysis could potentially detect code fragments that are not executed dynamically but contain anomalies related to heap or stack allocations, such as large loops or recursion. Combining dynamic and static research methods could be a subject for future research.

\textbf{How much did operating system implementation details affected our results?}
Heap allocations and deletions take varying amounts of time on different operating systems, and with different memory management algorithms and optimizers~\citep{10.1145/3373376.3378525,846583,hasan20061051}. Factors such as the size of objects and their distribution over time may also affect the performance of the |malloc()| and |free()| system calls. In this research we were using only one operating system (Ubuntu) of one specific version (22.04). This could be a subject for future research: determining which operating systems more quickly allocate C++ objects.

\textbf{What are limitations of our approach?}
The primary limitation of our study is that it relies on executable files without direct analysis of the source code. To maximize data collection, we focused exclusively on executables that do not require user input, encompassing mostly tests, as well as some benchmarks and examples. Consequently, it is not feasible to draw definitive conclusions about the overall code in the repositories, as the behavior of executables can differ significantly from how objects and functions are utilized in the source code of an application or library. Future studies could benefit from a detailed analysis of repository source code.

\section{Related Work}\label{sec:related}

We are not the first who are interested in C++ overheads. For example,
\citet{10.1145/236337.236369} measured the cost of virtual functions calls and found that C++ programs spend a median of 5.2\% of their time and 3.7\% of their instructions in dispatch code (overhead generated by virtual functions).

Previous studies have also delved into dynamic memory allocation in C++ programs. \citet{00122-9} put forward a hypothesis that dynamic memory management is presented explicitly and implicitly in constructors, copy constructors, overloading the assignment operator (\texttt{=}), type conversions, and application-specific member functions. They developed a source code level tracing tool to investigate the hypothesis. Similarly, \citet{485574} also studied memory allocations in object-oriented programs. They presented a hardware realization of the high-performance memory allocator for object-oriented systems. Additionally, \citet{846583} proposed an object reuse strategy to optimize memory allocations in object-oriented programs. They tested five C++ applications that can be obtained publicly. The results from all of them support their hypotheses that most objects are of sizes less than 512 bytes, short-lived, and highly reused. Thus, they developed a high-performance memory management strategy, a specialized dynamic memory allocator for reusable objects.

\citet{10.1002/spe.4380240602} chose five popular allocators at that time and measured their performance on eleven different C and C++ programs. They collected detailed statistics of the cost of dynamic storage allocation and analyzed it for each allocator.

To our best knowledge, there were no research as of yet measuring the popularity of different object allocation methods.

\section{Conclusion}\label{sec:conclusion}

In C++ programming, the choice of object allocation---whether in static memory, on the stack, or in the heap—has a significant impact on performance. This study investigated object allocation practices within 23 open-source C and C++ projects, focusing on the prevalence and implications of stack versus heap allocations during runtime. Utilizing DynamoRIO and Valgrind for empirical measurements, we found a vast range in the ratio of heap to stack allocations, from 0\% to 99\%. Contrary to the expectation that programmers might avoid heap allocations due to their high performance costs, our results indicate that heap allocations are extensively used in both C and C++ programs. Future research should explore the types of objects involved in these allocations to further understand allocation strategies.

\section{Data Availability}

All our scripts together with the data collected are published in Zenodo: \url{https://zenodo.org/records/11068271}.



\small
\bibliographystyle{plainnat}
\bibliography{main}

\begin{thebibliography}{12}
\providecommand{\natexlab}[1]{#1}
\providecommand{\url}[1]{\texttt{#1}}
\expandafter\ifx\csname urlstyle\endcsname\relax
  \providecommand{\doi}[1]{doi: #1}\else
  \providecommand{\doi}{doi: \begingroup \urlstyle{rm}\Url}\fi

\bibitem[Aigner and H{\"o}lzle(1996)]{aigner1996eliminating}
Gerald Aigner and Urs H{\"o}lzle.
\newblock {Eliminating Virtual Function Calls in C++ Programs}.
\newblock In \emph{{Proceedings of the European Conference on Object-Oriented
  Programming}}, pages 142--166. Springer, 1996.

\bibitem[Budimli{\'c} et~al.(1999)Budimli{\'c}, Kennedy, and
  Piper]{budimlic1999cost}
Zoran Budimli{\'c}, Ken Kennedy, and Jeff Piper.
\newblock {The Cost of Being Object-Oriented: A Preliminary Study}.
\newblock \emph{{Scientific Programming}}, 7\penalty0 (2):\penalty0 87--95,
  1999.

\bibitem[Chang et~al.(2001)Chang, Lee, and Srisa-an]{00122-9}
J.~Morris Chang, Woo~Hyong Lee, and Witawas Srisa-an.
\newblock {A Study of the Allocation Behavior of C++ Programs}.
\newblock \emph{Journal of Systems and Software}, 57\penalty0 (2):\penalty0
  107--118, 2001.
\newblock \doi{10.1016/S0164-1212(00)00122-9}.

\bibitem[Chang and Gehringer(1996)]{485574}
J.M. Chang and E.F. Gehringer.
\newblock {A High Performance Memory Allocator for Object-Oriented Systems}.
\newblock \emph{IEEE Transactions on Computers}, 45\penalty0 (3):\penalty0
  357--366, 1996.
\newblock \doi{10.1109/12.485574}.

\bibitem[Detlefs et~al.(1994)Detlefs, Dosser, and Zorn]{10.1002/spe.4380240602}
David Detlefs, Al~Dosser, and Benjamin Zorn.
\newblock {Memory Allocation Costs in Large C and C++ Programs}.
\newblock \emph{Software: Practice and Experience}, 24\penalty0 (6):\penalty0
  527--542, 1994.
\newblock \doi{10.1002/spe.4380240602}.

\bibitem[Driesen and H\"{o}lzle(1996)]{10.1145/236337.236369}
Karel Driesen and Urs H\"{o}lzle.
\newblock {The Direct Cost of Virtual Function Calls in C++}.
\newblock In \emph{{Proceedings of the 11th SIGPLAN Conference on
  Object-Oriented Programming, Systems, Languages, and Applications}}, pages
  306--323, 1996.
\newblock \doi{10.1145/236337.236369}.

\bibitem[Driesen and H{\"o}lzle(1996)]{driesen1996direct}
Karel Driesen and Urs H{\"o}lzle.
\newblock {The Direct Cost of Virtual Function Calls in C++}.
\newblock In \emph{{Proceedings of the Conference on Object-Oriented
  Programming, Systems, Languages, and Applications (OOPSLA)}}, pages 306--323,
  1996.

\bibitem[Hasan and Chang(2006)]{hasan20061051}
Yusuf Hasan and J.~Morris Chang.
\newblock {A Tunable Hybrid Memory Allocator}.
\newblock \emph{{Journal of Systems and Software}}, 79\penalty0 (8):\penalty0
  1051--1063, 2006.
\newblock \doi{10.1016/j.jss.2005.09.003}.

\bibitem[Lee et~al.(2000)Lee, Chang, and Hasan]{846583}
Woo~Hyong Lee, J.~M. Chang, and Y.~Hasan.
\newblock {Evaluation of a High-Performance Object Reuse Dynamic Memory
  Allocation Policy for C++ Programs}.
\newblock In \emph{{Proceedings of the 4th International Conference/Exhibition
  on High Performance Computing in the Asia-Pacific Region}}, volume~1, pages
  386--391, 2000.
\newblock \doi{10.1109/HPC.2000.846583}.

\bibitem[Maas et~al.(2020)Maas, Andersen, Isard, Javanmard, McKinley, and
  Raffel]{10.1145/3373376.3378525}
Martin Maas, David~G. Andersen, Michael Isard, Mohammad~Mahdi Javanmard,
  Kathryn~S. McKinley, and Colin Raffel.
\newblock {Learning-Based Memory Allocation for C++ Server Workloads}.
\newblock In \emph{{Proceedings of the Twenty-Fifth International Conference on
  Architectural Support for Programming Languages and Operating Systems}},
  pages 541--556, 2020.
\newblock \doi{10.1145/3373376.3378525}.

\bibitem[Stroustrup(1985)]{cpp}
Bjarne Stroustrup.
\newblock {The C++ Programming Language}, 1985.

\bibitem[Stroustrup(2004)]{10.1007/11535409_1}
Bjarne Stroustrup.
\newblock {Keynote Speech: Abstraction and the C++ Machine Model}.
\newblock In \emph{{Proceedings of the First International Conference on
  Embedded Software and Systems}}, pages 1--13, 2004.
\newblock \doi{10.1007/11535409_1}.

\end{thebibliography}


\end{document}